\documentclass[superscriptaddress,nofootinbib,notitlepage]{revtex4-1}
\usepackage{amssymb}
\usepackage{amsfonts}
\usepackage{amsmath}
\usepackage{graphicx}
\usepackage{dcolumn}
\usepackage{bm}
\usepackage[latin1]{inputenc}
\usepackage[dvips]{hyperref}

\begin{document}

\title{The frame of fixed stars in Relational Mechanics}\thanks{The final publication is available at
Springer via http://dx.doi.org/10.1007/s10701-016-0042-7}
\author{Rafael Ferraro}
\email[Member of Carrera del Investigador Cient\'{\i}fico (CONICET,
Argentina); ]{ferraro@iafe.uba.ar}\affiliation{Instituto de
Astronom\'\i a y F\'\i sica del Espacio (IAFE, CONICET-UBA), Casilla
de Correo 67, Sucursal 28, 1428 Buenos Aires, Argentina.}
\affiliation{Departamento de F\'\i sica, Facultad de Ciencias
Exactas y Naturales, Universidad de Buenos Aires, Ciudad
Universitaria, Pabell\'on I, 1428 Buenos Aires, Argentina.\vskip1cm}

\begin{abstract}
Relational mechanics is a gauge theory of classical mechanics whose
laws do not govern the motion of individual particles but the
evolution of the distances between particles. Its formulation gives
a satisfactory answer to Leibniz's and Mach's criticisms of Newton's
mechanics: relational mechanics does not rely on the idea of an
\textit{absolute space}. When describing the behavior of small
subsystems with respect to the so called ``fixed stars'', relational
mechanics basically agrees with Newtonian mechanics. However, those
subsystems having huge angular momenta will deviate from the
Newtonian behavior if they are described in the frame of fixed
stars. Such subsystems naturally belong to the field of astronomy;
they can be used to test the relational theory.
\end{abstract}

\maketitle

\tableofcontents

\section{Introduction}

\label{SI}

Relational mechanics is a reformulation of classical mechanics
leading to dynamical equations that are valid in any frame. By
extending the laws of mechanics to any frame, relational mechanics
abolishes Newton's \textit{absolute space}. So, no privileged
(\textit{inertial})\ frames exist in relational mechanics because
its dynamical equations obey an extended symmetry: instead of being
invariant just under the Galilean group (uniform translations of
frames and rigid rotations of axes) they are invariant under
arbitrary time-dependent changes of orthonormal frames. In the
language of field theory, the changes of orthonormal frames
constitute the \textit{gauge symmetry }of relational mechanics.
Notoriously, there are frames (particular \textit{gauge choices})
where relational mechanics makes contact with Newton's laws: in any
frame where the angular momentum of the universe vanishes and the
center of mass of the universe moves at a constant velocity, the
dynamical equations become the Newton's laws. Nonetheless, this kind
of privileged frames (we call them \textit{Newtonian frames}) is
determined not by an abstract entity like the absolute space but by
the set of particles constituting the entire universe. This means
that relational mechanics gives an answer to Mach's criticism of
Newtonian mechanics; relational mechanics is a \textit{Machian}
theory \cite{Mach}.

\bigskip

In spite that Newton's laws can be retrieved in relational mechanics, the
solutions to such equations deserves a careful understanding. In a Newtonian
frame we do obtain the Keplerian orbits for the motion of planets; however,
an individual evolution means nothing in relational mechanics. An individual
evolution can be completely distorted by a change of frame (change of
gauge). The individual positions and velocities are not \textit{observables}
but gauge dependent variables. In relational mechanics the observables
(i.e., physically meaningful gauge invariant magnitudes) are the distances
between particles and the angles between the straight lines joining pairs of
particles. Indeed, relational equations govern just the evolution of
observables magnitudes, since such evolutions are independent of the frame
choice. Nevertheless, a given evolution of distances can be described
through different individual evolutions in different frames (analogously, in
electromagnetism one can describe a given field configuration through
different gauge-related potentials). This means that the gauge symmetry
endows mechanics with the essential \textit{relational} feature claimed by
Leibniz in his correspondence with Clarke \cite{Alexander}.

\bigskip

So, going back to the Keplerian orbit of a planet, the statement that the
planet returns to the original position in a period is not physical, unless
it can be restated in terms of distances or angles. The planet position
means nothing in relational mechanics. Particle positions are mere
gauge-dependent mathematical tools to understand the relations between
particles. Instead, a physical statement should tell about the time elapsed
between successive alignments of the Sun-planet direction and some
\textquotedblleft fixed star\textquotedblright . Notice that, since the
Newtonian solution under consideration possesses non-null angular momentum,
the rest of the universe would have an opposite angular momentum to cancel
out the total angular momentum (as required in a Newtonian frame). Therefore
the time elapsed between successive alignments of the Sun-planet direction
and some \textquotedblleft fixed star\textquotedblright\ is actually smaller
than the Keplerian period, since the rest of the universe is
counter-rotating in a Newtonian frame. Therefore, in a\ frame of fixed stars
the planet go faster than expected; its velocity involves a non-Keplerian
contribution that is proportional to the orbital radius. Of course, this
effect is completely negligible in the case of a planetary system. In fact,
the angular momentum of a planetary system is compensated by an unnoticeable
rotation of the rest of the universe. Because of this reason, the
\textquotedblleft fixed stars\textquotedblright\ are really adequate to
establish an \textit{external} Newtonian frame for studying a planetary
system. But, what if we are studying a galaxy or a cluster of galaxies?

The rest of the article is organized as follows. In Section
\ref{SII} we give a brief account of the history of relational
mechanics. Besides we summarize the basics of relational mechanics
as they are developed in Ref.~\cite{Ferraro}. In Section \ref{SIII}
we show the dragging that affects the rotation curves of subsystems
displaying a huge angular momentum, when seen in the frame of fixed
stars. In Section \ref{SIV} we derive the relational virial theorem.
In Section \ref{SV} we study the relational two-bodies problem. In
Section \ref{SVI} we display the conclusions.

\bigskip

\section{Relational Mechanics in brief}

\label{SII} The idea of absolute space, as a way of designating the
privileged inertial frames where Newton's laws are valid, was criticized
from the very beginning of the science of mechanics. In Leibniz's opinion,
mechanics should describe \textit{relations} among bodies, rather than
individual evolutions relative to metaphysically defined frames \cite%
{Alexander}. Even though Newton was aware of this weakness of his
formulation --in the sense that the absolute motion cannot be evidenced--,
instead he thought that the absolute acceleration was a valid concept.
According to Newton, the absolute acceleration is evidenced by the parabolic
shape of water surface in a (absolutely) rotating bulk. However, Mach
objected this idea by stating that the shape of the water only proves the
rotation with respect to the rest of the universe \cite{Mach}, since nobody
knows what would happen if the water and the bulk were the only bodies in
the universe. Mach's criticism was a trigger in Einstein's route towards
general relativity. Einstein baptized \textquotedblleft Mach's
principle\textquotedblright\ the idea that inertia is determined by the
interaction with the rest of the universe \cite%
{Einstein12,Einstein14,Einstein16,Einstein18}.

The 20th century is rich in proposals to reformulate the mechanics starting
from relational principles. The laws of mechanics combine potentials, which
describe forces, and kinetic variables describing motion. The potentials are
already relational, since they just contain the distances $r_{ij}$ between
particles:%
\begin{equation}
V\ \ =\ \ \sum_{i<j}\ V_{ij}(r_{ij})\ =\ \ \frac{1}{2}\ \sum_{i\neq j}\
V_{ij}(r_{ij})\ ,
\end{equation}%
($V_{ij}=V_{ji}$). Instead, the Newtonian kinetic energy is made of
individual velocities; so it should be reformulated in terms of relative
velocities and, possibly, distances. Early attempts of this sort can be
found in References \cite{Hofmann,Reissner,Schrodinger,Barbour75} (for a
comprehensive account of these early tries see Ref.~\cite{Barbour95}).
However these attempts led to anisotropies of the inertia that are not
observationally supported \cite{Hughes,Barbour77}. After this setback it was
realized that the basic structure of the Newtonian kinetic energy should be
preserved in some sense in order to keep essential features of the
successful Newtonian mechanics. Noticeably, the form of the Newtonian
kinetic energy is strongly linked to the Galilean transformations, the
transformations between inertial frames. However, the aim of relational
mechanics is putting all the frames on an equal footing, with the consequent
abolition of the absolute space. For this, the Galilean \textit{rigid}
symmetry of Newton's theory should be extended to encompass any
time-depending translation and rotation,%
\begin{equation}
\mathbf{r}_{i}~\longrightarrow ~\mathbf{r}_{i}~+~\mathbf{\xi }(t)\ ,
\label{translation}
\end{equation}%
\begin{equation}
\mathbf{r}_{i}~\longrightarrow ~\mathbf{r}_{i}~+\mathbf{\alpha }(t)~\times ~%
\mathbf{r}_{i}~,  \label{rotation}
\end{equation}%
what in field theory is called \textit{gauging} the symmetry\footnote{$%
\mathbf{\alpha }(t)$ is an infinitesimal vector directed along the axis of
rotation (finite rotations require orthonormal matrices). Galileo
transformations are included in the gauge group (\ref{translation}), (\ref%
{rotation});\ they are the elements having ${\dot{\mathbf{\xi }}}=\mathbf{V}%
= $ constant, and $\dot{\mathbf{\alpha }}=0$.}. In a gauge theory
each \textit{physical state} is described by a set of
\textit{equivalent configurations}, all of them connected by gauge
transformations. In our case a physical state is determined for the
distances between particles, which can be read in terms of
equivalent configurations of individual positions in different
frames. In the language of gauge theory each set of equivalent
configurations is called \textit{orbit}, which represents a physical
state. While the Newtonian kinetic energy is related to a
\textit{measure} $\sum
\,m_{i}\,d\mathbf{r}_{i}\cdot d\mathbf{r}_{i}$ between near configurations $%
\left\{ \mathbf{r}_{i}\right\} $ and $\left\{ \mathbf{r}_{i}+d\mathbf{r}%
_{i}\right\} $, what is needed to build a relational kinetic energy is a
measure between near orbits. Such a measure will be automatically gauge
invariant. Not surprisingly, the measure between orbits can be obtained from
the (Newtonian) measure between configurations. This idea was developed in
Ref.~\cite{Barbour82}, where the measure between near orbits was defined as
the lower bound of the (Newtonian) measures between configurations
representative of each orbit. The measure between near orbits leads to a
gauge invariant kinetic energy, as is needed to formulate the relational
mechanics. This procedure is called \textit{best matching} \cite%
{Barbour03,Gryb,Anderson,Mercati}.

\bigskip

Another way to build a gauge invariant kinetic energy involves the
concept of \textit{covariant} derivative, as is typical in gauge
theory. When a rigid symmetry is gauged, the behavior of the
ordinary derivative under the so extended symmetry becomes
inappropriate (in a sense that will be explained below). The
covariant derivative includes a term to heal this undesirable
behavior. These strategy was followed in Ref.~\cite{Ferraro}, whose
results can be summarized as follows:

\bigskip

$\blacktriangleright$ The relational kinetic energy is built of relative
positions $\mathbf{r}_{i\,j}\doteq \mathbf{r}_{i}-\mathbf{r}_{j}$ and their
derivatives $\mathbf{v}_{i\,j}\doteq \dot{\mathbf{r}}_{i\,j}=\mathbf{v}_{i}-%
\mathbf{v}_{j}$. Both of them are invariant under time dependent
translations (\ref{translation}). However, even though $\mathbf{r}_{i\,j}$
behaves as a vector under time dependent rotations (\ref{rotation}), $%
\mathbf{v}_{i\,j}$ does not. In fact, from Eq.~(\ref{rotation}) it follows
that $\mathbf{r}_{i\,j}\longrightarrow \mathbf{r}_{i\,j}+\mathbf{\alpha }%
(t)\times \mathbf{r}_{i\,j}$ but%
\begin{equation}
\dot{\mathbf{r}}_{i\,j}~\longrightarrow ~\dot{\mathbf{r}}_{i\,j}~+~\mathbf{%
\alpha }\times \dot{\mathbf{r}}_{i\,j}~+~\dot{\mathbf{\alpha }}\times
\mathbf{r}_{i\,j}\ .  \label{badderivative}
\end{equation}%
So, the ordinary derivative of a vector does not behave as a vector under
time-depending rotations; the last term in Eq.~(\ref{badderivative}) must be
healed by means of the compensating mechanism of a covariant derivative.

\bigskip

$\blacktriangleright$ For an isolated system of particles
representing the entire universe, which is governed by classical
interactions at a distance that
are described by a potential $V$ depending on the distances $r_{i\,j}=|%
\mathbf{r}_{i\,j}|$, the compensating term in the covariant derivative (the
\textit{connection}) is built of the \textit{intrinsic} angular momentum $%
\mathbf{J}$ and inertia tensor $\mathbf{I}$\footnote{%
We call intrinsic those quantities of the form $\sum\limits_{i<j}\frac{%
m_{i}m_{j}}{2M}~f_{ij}(\mathbf{r}_{ij},\mathbf{v}_{ij})$ where $%
f_{ij}=f_{ji\,}$.}:%
\begin{equation}
\mathbf{J}~\ \doteq ~\ \sum\limits_{i<j}\,\frac{m_{i}~m_{j}}{M}~\mathbf{r}%
_{i\,j}\times \mathbf{v}_{i\,j}\ ,~  \label{J}
\end{equation}%
\begin{equation}
\mathbf{I}~\doteq ~\sum\limits_{i<j}\frac{m_{i}~m_{j}}{M}~[r_{i\,j}^{2}~%
\mathbf{1}-\mathbf{r}_{i\,j}\otimes \mathbf{r}_{i\,j}]\ \ .  \label{I}
\end{equation}%
In fact, since $\mathbf{J}$ contains relative velocities, then it
behaves as a vector just under rigid rotations. But under
time-dependent rotations $\delta \mathbf{J}$ gets a term
proportional to $\dot{\mathbf{\alpha }}$:
\begin{eqnarray}
\delta \mathbf{J}~\ &=&~\ \sum\limits_{i<j}\,\frac{m_{i}~m_{j}}{M}~\mathbf{r}%
_{i\,j}\times \delta \mathbf{v}_{i\,j}\ +\ ...\ =~...\ +\ \sum\limits_{i<j}\,%
\frac{m_{i}~m_{j}}{M}~\mathbf{r}_{i\,j}\times (~\dot{\mathbf{\alpha }}\times
\mathbf{r}_{i\,j})\ +\ ...  \notag \\
&&  \notag \\
&=&~...\ +\ \sum\limits_{i<j}\,\frac{m_{i}~m_{j}}{M}~\left[ \dot{\mathbf{%
\alpha }}\ r_{i\,j}^{2}-\mathbf{r}_{i\,j}\ \mathbf{r}_{i\,j}\cdot \dot{%
\mathbf{\alpha }}\right] \ +\ ...\ =~...\ +\ \mathbf{I}\cdot \dot{\mathbf{%
\alpha }}\ +\ ...\ .  \label{deltaJ}
\end{eqnarray}%
So $\mathbf{I}^{-1}\cdot \mathbf{J}$ is what is needed for canceling
out the last term in Eq.~(\ref{badderivative}). The
\textit{vectorial} relative velocity is defined as the covariant
derivative of the relative position,
\begin{equation}
\frac{D\mathbf{r}_{i\,j}}{Dt}\ \doteq \ \frac{d\mathbf{r}_{i\,j}}{dt}\ -\ (%
\mathbf{I}^{-1}\cdot \mathbf{J})\ \times \ \mathbf{r}_{i\,j~}\ ,
\label{covariant}
\end{equation}%
and the gauge invariant kinetic energy has the intrinsic form%
\begin{equation}
T~\ \doteq ~\ \sum\limits_{i<j}\,\frac{m_{i}~m_{j}}{2M}~\frac{D\mathbf{r}%
_{i\,j}}{Dt}\cdot \frac{D\mathbf{r}_{i\,j}}{Dt}\ .  \label{KE}
\end{equation}

$\blacktriangleright$ The gauge invariant kinetic energy (\ref{KE})\ can be
rephrased in several ways:
\begin{subequations}
\begin{eqnarray}
T~ &=&~\sum\limits_{i<j}\frac{m_{i}~m_{j}}{2M}~\mathbf{v}_{ij}\cdot \mathbf{v%
}_{ij}~-~\frac{1}{2}~\mathbf{J}\cdot \mathbf{I}^{-1}\cdot \mathbf{J}\ ,
\label{Ta} \\
&&  \notag \\
&=&~\sum\limits_{k}\frac{m_{k}}{2}~\left\vert \mathbf{v}_{k}-\frac{\mathbf{P}%
}{M}-(\mathbf{I}^{-1}\cdot \mathbf{J})\times (\mathbf{r}_{k}-\mathbf{R}%
)\right\vert ^{2}\ ,  \label{Tb} \\
&&  \notag \\
&=&~\sum\limits_{k}\frac{m_{k}}{2}~\left\vert \frac{D}{Dt}(\mathbf{r}_{k}-%
\mathbf{R})\right\vert ^{2}\ ,  \label{Tc}
\end{eqnarray}%
where $\mathbf{R}$ is the center-of-mass position, and $\mathbf{P}\doteq
\sum \,m_{k}\,\mathbf{v}_{k}$ is the total momentum ($\mathbf{R}$ and $%
\mathbf{P}$ are gauge dependent magnitudes). Equation~(\ref{Tc}) shows that
the Newtonian structure of the kinetic energy, as a sum of individual
particle contributions, has been preserved. However, $\mathbf{r}_{k}-\mathbf{%
R}$ replaces $\mathbf{r}_{k}$ to fulfill the invariance under time-dependent
translations, and the covariant derivative takes the role of the ordinary
derivative to fulfill the invariance under time-dependent rotations.

The kinetic energy (\ref{Ta}) originally appeared in References \cite%
{Lynden92,Lynden95,Katz95} where, instead of gauging the symmetry, the
authors obtained $T$ by means of the minimization of the Newtonian kinetic
energy with respect to translations and rotations, so making contact with
the best matching ansatz of Ref.~\cite{Barbour82}.

\bigskip

$\blacktriangleright$ The relational dynamical equations coming from the
Lagrangian $L(\mathbf{r}_{k},\,\mathbf{v}_{k})=T-V$ are
\end{subequations}
\begin{equation}
m_{k}~\frac{d}{dt}\left[ \mathbf{v}_{k}-\frac{\mathbf{P}}{M}-(\mathbf{I}%
^{-1}\cdot \mathbf{J})\times (\mathbf{r}_{k}-\mathbf{R})\right] ~=~-\mathbf{%
\nabla }_{k}\left( V+\frac{1}{2}\ \mathbf{J}\cdot \mathbf{I}^{-1}\cdot
\mathbf{J}\right) ~.  \label{motion}
\end{equation}%
On the l.h.s.~the expression inside brackets is $D(\mathbf{r}_{k}-\mathbf{R}%
)/Dt$. Since $d/dt$ is not a covariant derivative, then the
l.h.s.~is not a vector under gauge transformations; its bad behavior
is compensated by the (centrifugal) gauge-dependent term on the
r.h.s.

In spite of appearances, the equations (\ref{motion}) do not govern
the dynamics of $N$ individual particles. In fact, the number of
degrees of freedom is not $3N$ but $3N-6$ because the freedom to
choose the frame involves six parameters\footnote{See
Ref.~\cite{Ferraro} for the structure of constraints in the
Hamiltonian formulation of the theory.}. In other words, the
evolutions are determined modulo arbitrary time-dependent
translations and rotations, because of the gauge invariance
displayed by the Lagrangian and the dynamical equations. For
instance, a configuration where the system rigidly rotates (i.e.,
the distances between particles remain unchanged) is equivalent to
the configuration of \textit{rest}. The $3N-6$ degrees of freedom
can be associated with the minimum number of distances that are
needed to describe a state. For $N=2$ particles, just one distance
is involved (since the rotation around the axis of symmetry is
meaningless, the system has $3N-5=1$ degrees of freedom). For $N=3$
the particles form a triangle described by $3$ distances or degrees
of freedom. For $N>3$, each added particle must come at least with a
\textquotedblleft tripod\textquotedblright\ of $3$ distances to
determine its position with respect to the other particles. Thus,
the minimum number of distances to describe a state is
$3+3(N-3)=3N-6$, ($N>2$). Therefore, the equations (\ref{motion})
govern a dynamics of distances
instead of individual evolutions. As a consequence, the equations (\ref%
{motion}) remain strongly coupled even in the absence of interaction. In
fact, all the particles of the system are contained in $\mathbf{R}$, $%
\mathbf{P}$, $\mathbf{J}$ and $\mathbf{I}$. Nevertheless, the gauge
dependent magnitudes $\mathbf{P}$, $\mathbf{J}$ can be fixed by
choosing an appropriate frame where $\mathbf{P}$ is constant and
$\mathbf{J}$ vanishes. In such \textit{Newtonian frames}, that are
determined by the entire
distribution of mass in the universe, the equations of motion become%
\begin{equation}
m_{k}~\frac{d\mathbf{v}_{k}}{dt}~=~-\mathbf{\nabla }_{k}V~.  \label{Newton}
\end{equation}%
These gauge-fixed dynamical equations could create the illusion that
Newton's dynamics is got at the end of the day, since a set of
individual evolutions fulfilling the Newton's laws has been
obtained. However, we must keep in mind that the individual
evolutions means nothing in relational mechanics. They are not
observables at all but gauge-dependent magnitudes. The observables
are the distances between particles. So, even if we solve the
equations (\ref{Newton}), we must analyze the meaning of such
Newtonian solutions in terms of relative distances (or angles).
Actually the equations (\ref{Newton}) must be solved together with
the gauge conditions $\mathbf{P}=constant$ and $\mathbf{J}=0$. Any
Newtonian evolution (of the entire universe) makes sense if and only
if the gauge conditions are fulfilled as well.

\section{Interpretation of Newtonian solutions}

\label{SIII}
\begin{figure}[t]
\centering \includegraphics[width=7.3cm]{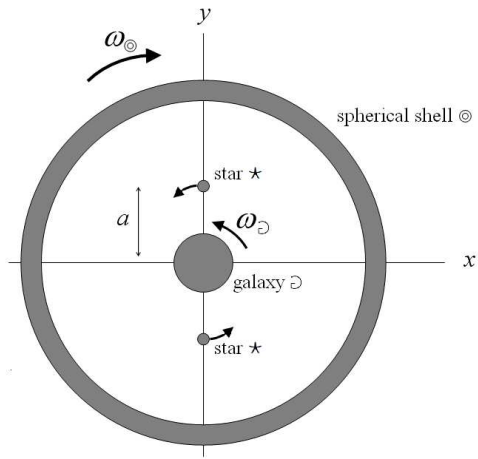} \caption{Two
stars orbiting a galaxy. The counter-rotating shell represents the
rest of the universe in the Newtonian frame where the total
intrinsic angular momentum vanishes.} \label{Fig1}
\end{figure}

Let us consider the idealized situation of a subsystem
gravitationally isolated from the rest of the island universe. Figure \ref%
{Fig1} shows a self-gravitating subsystem composed by two equal point-like
objects of mass $m$ (the stars $\star $) sharing a circular orbit of radius $%
a$\ around a central object (the galaxy $\Game $), surrounded by a rigid
isotropic spherical shell (representing the rest of the universe $%
\circledcirc $). The shell has not gravitational influence on the subsystem
inside it. According to Newton's second law (\ref{Newton}), the time the
stars complete their circular orbit is the Keplerian period%
\begin{equation}
\tau _{Kepler}\ =\ 2\pi \ \sqrt{\frac{a^{3}}{G\ (M_{\Game }+\frac{m_{\star }%
}{4})}}\   \label{TKepler}
\end{equation}%
(the contribution $m_{\star }/4$ comes from the gravitational
interaction between the stars). Furthermore, in Newtonian mechanics
the orbital solution exists irrespective of the presence of the
central object and the shell. Instead, in relational mechanics any
Newtonian solution comes together with the gauge condition
$\mathbf{J}=0$. Since we have split the universe into three parts
sharing their centers-of-mass, then the vanishing of the total
intrinsic angular momentum reads\footnote{\label{fn} It is easy to
verify that the intrinsic magnitudes $\mathbf{J}$ and $\mathbf{I}$
are the usual
angular momentum and tensor of inertia with respect to the center of mass: $%
\mathbf{J}=\sum \,m_{k}~(\mathbf{r}_{k}-\mathbf{R})\times \mathbf{v}_{k}~$, $%
\mathbf{I}=\sum \,m_{k}\,[|\mathbf{r}_{k}-\mathbf{R}|^{2}~\mathbf{1}-(%
\mathbf{r}_{k}-\mathbf{R})\otimes (\mathbf{r}_{k}-\mathbf{R})]$. In
general, the intrinsic magnitudes are not additive; the intrinsic
angular momentum of the universe is not the sum of the intrinsic
angular momentum (spin) of its parts because of orbital
contributions. However, if the system is split into several parts
whose centers-of-mass are coincident, as in the case of Figure
\ref{Fig1}, then $\mathbf{J}$ and $\mathbf{I}$ can be decomposed as
the sum of the intrinsic quantities belonging to each part, as done in Eq.~(%
\ref{J=0}). In general, if the system is split into two parts $A$ and $B$,
then it follows that $\mathbf{J}=\mathbf{J}_A+\mathbf{J}_B+(\mathbf{R}_A-%
\mathbf{R}_B)\times(M_B\,\mathbf{P}_A-M_A\,\mathbf{P}_B)/M$.}
\begin{equation}
\mathbf{J}_{\star }\ +\ \mathbf{J}_{\Game }\ +\ \mathbf{J}_{\circledcirc }\
=0\ ,\   \label{J=0}
\end{equation}%
where $\mathbf{J}_{\star }$, $\mathbf{J}_{\Game }$, and $\mathbf{J}%
_{\circledcirc }$ are the intrinsic angular momentum of each part
(the rigid shell rotates at a constant velocity, as required by the
equations (\ref{Newton})). Thus the presences of the central body or
the shell are essential to accomplish the requirement (\ref{J=0}),
so making sense to a Newtonian solution that displays a non-null
angular momentum $\mathbf{J}_{\star}$\footnote{A system of just
$N=2$ particles has only one degree of freedom (the distance between
the particles). The circular motion would imply that the distance is
constant. But this would only be possible in the absence of
interaction. The role played by the rest of the universe as
responsible of the centrifugal effect that is needed to sustain the
orbital motion (and the
shape of the water in the Newton's bucket as well) is analyzed in Ref.~\cite%
{Ferraro}.}.

Let us now focus on the meaning of $\tau _{Kepler}$ in relational
mechanics. While $\tau _{Kepler}$ in Newtonian mechanics is the
period of the circular motion in the privileged inertial frames,
$\tau _{Kepler}$ in relational mechanics is the period between equal
positions on the circle, as seen in a particular gauge fixed frame.
However positions are not observables but gauge dependent
magnitudes. The circular motion would look completely distorted in
an (equally allowed) arbitrarily rotating frame. A physical (gauge
independent) interval of time should allude only to observables. The
obervables in relational mechanics are the distances between
particles and the angles between lines joining particles. The
statement that the orbit is circular is gauge invariant, because is
based on the constancy of a distance. Instead, the time elapsed
between successive passes through the same position of a circular
orbit is a gauge dependent concept. So, to introduce a period $\tau$
defined in terms of observables, we should resort to the successive
passes of the stars through the line joining the center of the
circular orbit to some \textquotedblleft fixed
star\textquotedblright\ in the shell. According to Eq.~(\ref{J=0}),
the shell is counter-rotating with respect to the stars-galaxy
subsystem. If the stars co-rotate with the galaxy, then the interval
$\tau $ between successive passes through the line
is lower than $\tau _{Kepler}$:%
\begin{equation}
\mathbf{\omega }_{\star }\,\tau \ =\ \mathbf{\omega }_{\circledcirc }\,\tau
\ +\ 2\pi \,\,\widehat{\mathbf{z}}\ \ ,
\end{equation}%
where $\mathbf{\omega }_{\star }=(2\pi /\tau _{Kepler})\,\widehat{\mathbf{z}}
$ and $\mathbf{\omega }_{\circledcirc }=\mathbf{J}_{\circledcirc
}/I_{\circledcirc }=-(2\ \mathbf{J}_{\star }+\mathbf{J}_{\Game
})/I_{\circledcirc }\simeq -J_{\Game }/I_{\circledcirc }\,\widehat{\mathbf{z}%
}$. Then, it follows that the time $\tau $ is\footnote{In an
elliptic orbit, however, $\tau _{Kepler}$ is the time elapsed
between successive passes through the periastron.\ This is an
observable, since the periastron is defined by the minimization of a
distance. Notoriously the periastron suffers a cumulative shift in
the frame of fixed stars because $\tau _{Kepler}>\tau $.}
\medskip
\begin{equation}
\tau \ \simeq \ \frac{\tau _{Kepler}}{1+\frac{\tau _{Kepler}\ J_{\Game }}{%
2\pi \,\,I_{\circledcirc }}}\ .\   \label{tau}
\end{equation}%
So, the speed $v^{\prime }=2\pi \,\,a/\tau $ contains a
non-Keplerian contribution:
\begin{equation}
v^{\prime }\ =\ \frac{2\pi \,\,a}{\tau }\ = \ \ \frac{2\pi
\,\,a}{\tau _{Kepler}}\ +\ \frac{a\ J_{\Game }}{\,I_{\circledcirc
}}\ =\ \ v_{Kepler}\ +\ \frac{a\ J_{\Game }}{\,I_{\circledcirc }}\ .
\label{Coriolis}
\end{equation}%
The last term is a typical Coriolis effect. In fact, $v^{\prime }$ in Eq.~(%
\ref{Coriolis}) is the velocity in the frame of \textquotedblleft fixed
stars\textquotedblright , which is a non-Newtonian frame. In a non-Newtonian
frame, the Eq.~(\ref{motion}) contains not only the interaction forces
deriving from $V$ but contributions associated with the mean rotation of the
universe $\mathbf{I}^{-1}\cdot \mathbf{J}$ and the acceleration $\dot{%
\mathbf{P}}/M$. These extra contributions have the form of the
inertial forces of Newtonian mechanics \cite{Ferraro}, although they
are not determined by the absolute space but by the distribution of
matter in the universe. Then, there exists a Coriolis effect on the
velocities measured in the frame of fixed stars\footnote{The
measurement of velocities in the universe involves the Doppler
shift. Ignoring general relativity effects that are beyond this
framework, the Doppler shift depends on the relative radial velocity
source-observer, which is a gauge invariant magnitude (it is the
change of a distance per unit of time).}.

As a part of our Newtonian prejudices, we are used to accept that the axes
of an \textquotedblleft inertial\textquotedblright\ frame are pointed to
\textquotedblleft fixed stars\textquotedblright . This misconception seems
to be justified by experimental evidence at the scale of the solar system,
where Newton's laws work very well in the frame of fixed stars. However, a
frame of fixed stars could be a good approximation to a Newtonian frame only
to study subsystems of negligible angular momentum; this approximation would
fail when huge structures are considered.

We remark that $\mathbf{J}_{\Game }$ in Eqs.~(\ref{J=0}), (\ref{tau}) and (%
\ref{Coriolis}) is the angular momentum of the galaxy in the Newtonian
frame, which transforms to the frame of fixed stars as%
\begin{equation}
\mathbf{J}_{\Game }^{\prime }\ =\ \mathbf{J}_{\Game }-I_{\Game }\ \mathbf{%
\omega }_{\circledcirc }\ \simeq \ \left( 1+\frac{I_{\Game }}{%
I_{\circledcirc }}\right) \ \mathbf{J}_{\Game }\ .  \label{Jprime}
\end{equation}

\section{\protect\bigskip The relational virial theorem}

\label{SIV}By combining the equations of motion (\ref{motion}) we obtain
equations for $\mathbf{r}_{ij}$:%
\begin{equation}
m_{i}~m_{j}~\frac{d}{dt}\left[ \frac{D\mathbf{r}_{ij}}{Dt}\right] ~=~-(m_{j}~%
\mathbf{\nabla }_{i}-m_{i}~\mathbf{\nabla }_{j})\left( V+\frac{1}{2}\
\mathbf{J}\cdot \mathbf{I}^{-1}\cdot \mathbf{J}\right) ~.  \label{1}
\end{equation}%
The \bigskip intrinsic virial is a gauge invariant magnitude which can be
defined for any subsystem $\mathcal{S}$:%
\begin{equation}
G~\doteq ~\sum\limits_{i<j}^{N_{\mathcal{S}}}\,\frac{m_{i}~m_{j}}{2\,M_{%
\mathcal{S}}}~\frac{d}{dt}(r_{ij}^{2})~=~\sum\limits_{i<j}^{N_{\mathcal{S}%
}}\,\frac{m_{i}~m_{j}}{M_{\mathcal{S}}}~\mathbf{r}_{ij}\cdot \mathbf{v}%
_{ij}~=~\sum\limits_{i<j}^{N_{\mathcal{S}}}\,\frac{m_{i}~m_{j}}{M_{\mathcal{S%
}}}~\mathbf{r}_{ij}\cdot \frac{D\mathbf{r}_{ij}}{Dt}\ .
\end{equation}%
Its temporal derivative is%
\begin{eqnarray}
\frac{dG}{dt}~ &=&~\sum\limits_{i<j}^{N_{\mathcal{S}}}\,\frac{m_{i}~m_{j}}{%
M_{\mathcal{S}}}~\left( \mathbf{v}_{ij}\cdot \frac{D\mathbf{r}_{ij}}{Dt}+%
\mathbf{r}_{ij}\cdot \frac{d}{dt}\left[ \frac{D\mathbf{r}_{ij}}{Dt}\right]
\right) ~~  \notag \\
&&  \notag \\
&=&~\sum\limits_{i<j}^{N_{\mathcal{S}}}\,\frac{m_{i}~m_{j}}{M_{\mathcal{S}}}%
~\left( \mathbf{v}_{ij}\cdot \mathbf{v}_{ij}-\mathbf{v}_{ij}\cdot \ [(%
\mathbf{I}^{-1}\cdot \mathbf{J})\ \times \ \mathbf{r}_{i\,j~}]+\mathbf{r}%
_{ij}\cdot \frac{d}{dt}\left[ \frac{D\mathbf{r}_{ij}}{Dt}\right] \right)
\label{dG} \\
&&  \notag \\
&=&\ \sum\limits_{k=1}^{N_{\mathcal{S}}}\,m_{k}~\left( \mathbf{v}_{k}-\frac{%
\mathbf{P}_{\mathcal{S}}}{M_{\mathcal{S}}}\right) \cdot \left( \mathbf{v}%
_{k}-\frac{\mathbf{P}_{\mathcal{S}}}{M_{\mathcal{S}}}\right) ~-\mathbf{J}_{%
\mathcal{S}}\cdot \ \mathbf{I}^{-1}\cdot \mathbf{J}-~\sum\limits_{i<j}^{N_{%
\mathcal{S}}}\,\frac{1}{M_{\mathcal{S}}}~\mathbf{r}_{ij}\cdot (m_{j}~\mathbf{%
\nabla }_{i}-m_{i}~\mathbf{\nabla }_{j})\left( V+\frac{1}{2}\ \mathbf{J}%
\cdot \mathbf{I}^{-1}\cdot \mathbf{J}\right)  \notag
\end{eqnarray}%
(to get the first term, replace $\sum\nolimits_{i<j}$ with
$(1/2)\sum\nolimits_{i\neq \,j}$ and perform one of the involved
sums; the second term results from the circular shift property of
the mixed product). We will separately analyze the contributions of
$V$ and $\mathbf{J}$ to the last term of Eq.~(\ref{dG}).

\subsection{\protect\bigskip The potential $V$}

The role of the potential in $dG/dt$ does not differ from the respective one
in Newton's theory. In fact, by replacing $\sum\nolimits_{i<j}$ with $%
(1/2)\sum\nolimits_{i\neq \,j}$ we notice that both terms $m_{j}\mathbf{%
\nabla }_{i}$ and $-m_{i}\mathbf{\nabla }_{j}$ will make the same
contribution to the sum (to prove it, change $i\longleftrightarrow j$). So
it follows that

\begin{equation}
\frac{1}{2}\sum\limits_{i\neq j}^{N_{\mathcal{S}}}\,~\mathbf{r}_{ij}\cdot
\left( \frac{m_{j}}{M_{\mathcal{S}}}\mathbf{\nabla }_{i}-\frac{m_{i}}{M_{%
\mathcal{S}}}\mathbf{\nabla }_{j}\right) V\ =\ \sum\limits_{i\neq j}^{N_{%
\mathcal{S}}}\,~\frac{m_{j}}{M_{\mathcal{S}}}\ \mathbf{r}_{ij}\cdot \mathbf{%
\nabla }_{i}V\ =\ \sum\limits_{i=1}^{N_{\mathcal{S}}}\,~(\mathbf{r}_{i}-%
\mathbf{R}_{\mathcal{S}})\cdot \mathbf{\nabla }_{i}V~.
\end{equation}%
If the subsystem is isolated (only internal forces act on its particles)
then the total force is zero: $\sum_{i=1}^{N_{\mathcal{S}}}\mathbf{\nabla }%
_{i}V=0$. Therefore, the Newtonian result is obtained:%
\begin{equation}
\frac{1}{2}\sum\limits_{i\neq j}^{N_{\mathcal{S}}}\,~\mathbf{r}_{ij}\cdot
\left( \frac{m_{j}}{M_{\mathcal{S}}}\mathbf{\nabla }_{i}-\frac{m_{i}}{M_{%
\mathcal{S}}}\mathbf{\nabla }_{j}\right) V\ \ =\ \sum\limits_{i=1}^{N_{%
\mathcal{S}}}\,~\mathbf{r}_{i}\cdot \mathbf{\nabla }_{i}V~.
\end{equation}%
Since $V=\sum\nolimits_{i<j}^{N_{\mathcal{S}}}V_{ij}(r_{ij})$, it follows
that
\begin{equation}
\sum\limits_{i=1}^{N_{\mathcal{S}}}\,~\mathbf{r}_{i}\cdot \mathbf{\nabla }%
_{i}V\ =\ \sum\limits_{i=1}^{N_{\mathcal{S}}}\,~\mathbf{r}_{i}\cdot
\sum\limits_{j=1}^{N_{\mathcal{S}}}\frac{\partial V_{ij}}{\partial r_{ij}}%
\frac{\mathbf{r}_{ij}}{r_{ij}}\ =\ \frac{1}{2}\sum\limits_{i\neq j}^{N_{%
\mathcal{S}}}\,~\mathbf{r}_{ij}\cdot \frac{\partial V_{ij}}{\partial r_{ij}}%
\frac{\mathbf{r}_{ij}}{r_{ij}}\ =\ \frac{1}{2}\sum\limits_{i\neq j}^{N_{%
\mathcal{S}}}\,~r_{ij}~\frac{\partial V_{ij}}{\partial r_{ij}}\ =\
\sum\limits_{i<j}^{N_{\mathcal{S}}}\,~r_{ij}~\frac{\partial V_{ij}}{\partial
r_{ij}}\ .
\end{equation}%
If $V_{ij}\propto r_{ij}^{\alpha }$, then we obtain%
\begin{equation}
\frac{1}{2}\sum\limits_{i\neq j}^{N_{\mathcal{S}}}\,~\mathbf{r}_{ij}\cdot
\left( \frac{m_{j}}{M_{\mathcal{S}}}\mathbf{\nabla }_{i}-\frac{m_{i}}{M_{%
\mathcal{S}}}\mathbf{\nabla }_{j}\right) V\ \ =\ \alpha \ V_{\mathcal{S}}\ ,
\end{equation}%
where $V_{\mathcal{S}}$ is the internal potential energy of the isolated
subsystem.

\subsection{\protect\bigskip The total angular momentum $\mathbf{J}$}

It is fulfilled that \cite{Ferraro}%
\begin{equation}
\mathbf{\nabla }_{i}\left( \frac{1}{2}\ \mathbf{J}\cdot \mathbf{I}^{-1}\cdot
\mathbf{J}\right) ~=~-m_{i}\ (\mathbf{I}^{-1}\cdot \mathbf{J})\times \left[
\left( \mathbf{v}_{i}-\frac{\mathbf{P}}{M}\right) \ -(\mathbf{I}^{-1}\cdot
\mathbf{J})\times (\mathbf{r}_{i}-\mathbf{R})\right] \ ;
\end{equation}%
then, the last term in Eq.~(\ref{dG}) is%
\begin{eqnarray}
\mathbf{r}_{ij} &\cdot &\left( \frac{m_{j}}{M_{\mathcal{S}}}\mathbf{\nabla }%
_{i}-\frac{m_{i}}{M_{\mathcal{S}}}\mathbf{\nabla }_{j}\right) \left( \frac{1%
}{2}\ \mathbf{J}\cdot \mathbf{I}^{-1}\cdot \mathbf{J}\right) ~=~-\frac{%
m_{i}\ m_{j}}{M_{\mathcal{S}}}~\mathbf{r}_{ij}\cdot \left[ (\mathbf{I}%
^{-1}\cdot \mathbf{J})\times \left[ \mathbf{v}_{ij}\ -(\mathbf{I}^{-1}\cdot
\mathbf{J})\times \mathbf{r}_{ij}\right] \right] ~  \notag \\
&&  \notag \\
&=&~\frac{m_{i}\ m_{j}}{M_{\mathcal{S}}}~\left\{ (\mathbf{I}^{-1}\cdot
\mathbf{J})\cdot \left[ \mathbf{r}_{ij}\times \mathbf{v}_{ij}\right] ~~-~(%
\mathbf{I}^{-1}\cdot \mathbf{J})\cdot \left[ \mathbf{r}_{ij}\times \left[ (%
\mathbf{I}^{-1}\cdot \mathbf{J})\times \mathbf{r}_{ij}\right] \right]
\right\}  \notag \\
&&  \notag \\
&=&~\frac{m_{i}\ m_{j}}{M_{\mathcal{S}}}~\left\{ (\mathbf{I}^{-1}\cdot
\mathbf{J})\cdot \left[ \mathbf{r}_{ij}\times \mathbf{v}_{ij}\right] ~-~(%
\mathbf{I}^{-1}\cdot \mathbf{J})\cdot \left[ r_{ij}^{2}\ (\mathbf{I}%
^{-1}\cdot \mathbf{J})-\mathbf{r}_{ij}\ [\mathbf{r}_{ij}\cdot (\mathbf{I}%
^{-1}\cdot \mathbf{J})]\right] \right\} ~,
\end{eqnarray}%
which adds to%
\begin{equation}
\sum\limits_{i<j}^{N_{\mathcal{S}}}\,~\mathbf{r}_{ij}\cdot \left( \frac{m_{j}%
}{M_{\mathcal{S}}}\mathbf{\nabla }_{i}-\frac{m_{i}}{M_{\mathcal{S}}}\mathbf{%
\nabla }_{j}\right) \left( \frac{1}{2}\ \mathbf{J}\cdot \mathbf{I}^{-1}\cdot
\mathbf{J}\right) ~=~\mathbf{J}_{\mathcal{S}}\cdot \ \mathbf{I}^{-1}\cdot
\mathbf{J\,}-\,(\mathbf{I}^{-1}\cdot \mathbf{J})\cdot \mathbf{I}_{\mathcal{S}%
}\cdot (\mathbf{I}^{-1}\cdot \mathbf{J})~.
\end{equation}

\bigskip

\subsection{\protect\bigskip Summary and results}

In sum, the result is%
\begin{equation}
\frac{dG}{dt}~~=\ \sum\limits_{k=1}^{N_{\mathcal{S}}}\,m_{k}~\left( \mathbf{v%
}_{k}-\frac{\mathbf{P}_{\mathcal{S}}}{M_{\mathcal{S}}}\right) \cdot \left(
\mathbf{v}_{k}-\frac{\mathbf{P}_{\mathcal{S}}}{M_{\mathcal{S}}}\right) ~-2~%
\mathbf{J}_{\mathcal{S}}\cdot \ \mathbf{I}^{-1}\cdot \mathbf{J}+(\mathbf{I}%
^{-1}\cdot \mathbf{J})\cdot \mathbf{I}_{\mathcal{S}}\cdot (\mathbf{I}%
^{-1}\cdot \mathbf{J})-\alpha \ V_{\mathcal{S}}\ ,
\end{equation}%
where the first term is twice the Newtonian kinetic energy $T_{\mathcal{S}%
}^{Newton}$ in a frame where the subsystem-center-of-mass is at rest. The
virial theorem is based on the assumption that%
\begin{equation}
\frac{1}{\tau }\ \int_{0}^{\tau }\ \frac{dG}{dt}\ dt\ =\ \frac{G(\tau )-G(0)%
}{\tau }
\end{equation}%
is a vanishing quantity. This can happens both if the subsystem is periodic
of period $\tau $ (the distances $r_{ij}$ and their derivatives periodically
repeat themselves) or if the subsystem remains bounded for a (going to
infinity) very large time $\tau $. In such cases it follows that%
\begin{equation}
<2\ T_{\mathcal{S}}^{Newton}\ -2~\mathbf{J}_{\mathcal{S}}\cdot \ \mathbf{I}%
^{-1}\cdot \mathbf{J}+\ (\mathbf{I}^{-1}\cdot \mathbf{J})\cdot \mathbf{I}_{%
\mathcal{S}}\cdot (\mathbf{I}^{-1}\cdot \mathbf{J})>\ =\ \alpha \ <V_{%
\mathcal{S}}>\ .  \label{VT}
\end{equation}%
It is easy to check that the quantity inside the l.h.s.~brackets is twice
the gauge invariant kinetic energy of the subsystem (replace $%
\sum\nolimits_{i<j}$ with $\sum\nolimits_{i<j}^{N_{\mathcal{S}}}$ in Eq.~(%
\ref{KE})). The result (\ref{VT}) shows the influence of the rest of
the universe on the evolution of an \textquotedblleft
isolated\textquotedblright\ subsystem in an arbitrary frame. In a
Newtonian frame it is $\mathbf{J}=0$, so the Newtonian form of the
virial theorem is recovered. In the simple model where the rest of
the universe is represented by a rigid isotropic spherical shell,
the frame of fixed stars (i.e., fixed shell) is not strictly
Newtonian because a non-vanishing angular momentum still remains:
the angular momentum of the considered subsystem. In such a case it
is $\mathbf{J}^{\prime}=\mathbf{J}_{\mathcal{S}}^{\prime}$;
then, by replacing $\mathbf{I}_{\mathcal{S}}=\mathbf{I}-\mathbf{I}%
_{\circledcirc }$ (see Footnote \ref{fn}) we obtain the virial theorem in
the frame of fixed stars:
\begin{equation}
<2\ T_{\mathcal{S}}^{{\prime}\,Newton}\ -2~\mathbf{J}_{\mathcal{S}%
}^{\prime}\cdot \ \mathbf{I}^{-1}\cdot \mathbf{J}_{\mathcal{S}%
}^{\prime}+\ (\mathbf{I}^{-1}\cdot
\mathbf{J}_{\mathcal{S}}^{\prime})\cdot
(\mathbf{I}-\mathbf{I}_{\circledcirc })\cdot (\mathbf{I}^{-1}\cdot
\mathbf{J}_{\mathcal{S}}^{\prime})>\ =\ \alpha \ <V_{\mathcal{S}}>\
,
\end{equation}%
i.e.,%
\begin{equation}
<2\ T_{\mathcal{S}}^{{\prime}\,Newton}\
-\mathbf{J}_{\mathcal{S}}^{^{\prime
}}\cdot \ \mathbf{I}^{-1}\cdot \mathbf{J}_{\mathcal{S}}^{\prime}-\ (%
\mathbf{I}^{-1}\cdot \mathbf{J}_{\mathcal{S}}^{\prime})\cdot \mathbf{I}%
_{\circledcirc }\cdot (\mathbf{I}^{-1}\cdot \mathbf{J}_{\mathcal{S}%
}^{\prime})>\ =\ \alpha \ <V_{\mathcal{S}}>\ .
\end{equation}%
At the lowest level of approximation it is $\mathbf{I}\sim
\mathbf{I}_{\circledcirc }$; thus
\begin{equation}
<T_{\mathcal{S}}^{{\prime}\,Newton}>\, -\, <\mathbf{J}_{\mathcal{S}%
}^{\prime}\cdot \ \mathbf{I}_{\circledcirc }^{-1}\cdot \mathbf{J}_{%
\mathcal{S}}^{\prime}>\ \simeq \ \frac{\alpha}{2} \
<V_{\mathcal{S}}>\ .\label{approxvirial}
\end{equation}%
The relational virial theorem in a frame of fixed stars departs from
its Newtonian version depending on how large the subsystem intrinsic
angular momentum is.

\bigskip

If the subsystem $\mathcal{S}$ decomposes into two parts $\Game $ and $\star
$ with coincident centers-of-mass, then it follows that $\mathbf{J}_{%
\mathcal{S}}=\mathbf{J}_{\Game }+\mathbf{J}_{\star }$ and $\mathbf{I}_{%
\mathcal{S}}=\mathbf{I}_{\Game }+\mathbf{I}_{\star }$ (see Footnote \ref{fn}%
); besides it is $V_{\mathcal{S}}=V_{\Game }+V_{\star }+V_{\Game \star }$.
To exemplify, we will apply the virial theorem to the case studied in the
previous Section. Since the galaxy can be considered itself as an isolated
subsystem --the interaction with the stars is not relevant to its
evolution-- then an equation like (\ref{VT}) is separately valid for the
galaxy as well. Thus we are left with the following relation for the stars:%
\begin{equation}
<2\ T_{\star }^{Newton}\ -2~\mathbf{J}_{\star }\cdot \ \mathbf{I}^{-1}\cdot
\mathbf{J}+\ (\mathbf{I}^{-1}\cdot \mathbf{J})\cdot \mathbf{I}_{\star }\cdot
(\mathbf{I}^{-1}\cdot \mathbf{J})>\ =\ \alpha \ <V_{\star }+V_{\Game \star
}>\ .  \label{VTstars}
\end{equation}%
In the frame of fixed stars it is $\mathbf{J}^{\prime }=\mathbf{J}_{\mathcal{%
S}}^{\prime }=\mathbf{J}_{\Game }^{\prime }+\mathbf{J}_{\star }^{\prime
}\simeq J_{\Game }^{\prime }\,\,\widehat{\mathbf{z}}$. Therefore, it results%
\begin{equation}
2\ m\ v^{\prime 2}-2\ (2\ m\ v^{\prime }\ a)\ I^{-1}\ J^{\prime }\
+\ (2\ m\ a^{2})(I^{-1}\ J^{\prime })^{2}\ \ \simeq\ \ 2\ \frac{G\
M_{\Game }\ m}{a}\ ,
\end{equation}%
where $V_{\star }$ has been neglected. Thus%
\begin{equation}
(v^{\prime }-a\ I^{-1}\ J^{\prime })^{2}\ \simeq\ \frac{G\ M_{\Game
}}{a}\ =\ v_{Kepler}^{2}\ ,
\end{equation}%
i.e.,%
\begin{equation}
v^{\prime }\ \simeq\ v_{Kepler}\ +\ a\ I^{-1}\ J^{\prime }\ \simeq \
\ v_{Kepler}\ +\ \frac{a\ J_{\Game }}{\,I_{\circledcirc }}
\end{equation}%
in agreement with the result (\ref{Coriolis}).

\section{\protect\bigskip The relational two-body problem}

\label{SV}As shown in Section \ref{SIII}, the dynamics of an
\textquotedblleft isolated\textquotedblright\ two-body system in the frame
of fixed stars involves the constant $I_{\circledcirc }$, which appears as a
sort of universal constant. Let us consider two bodies $m_{1}$ and $m_{2}$
interacting through a potential $V(r_{12})$ much larger than the
interactions with the other particles of the universe. Like in Figure \ref%
{Fig1}, we will idealize the rest of the universe as a spherical shell
centered at the center-of-mass of the two-body system. Thus, no
gravitational field remains inside the shell apart from the interaction $%
V(r_{12})$. As explained in Footnote \ref{fn}, such configuration of
subsystems with a common center-of-mass implies additivity: $\mathbf{J}=%
\mathbf{J}_{\circledcirc }+\mathbf{J}_{12}$, where $\mathbf{J}_{\circledcirc
}$ stands for the intrinsic angular momentum of the rest of the universe
(for $\mathbf{J}_{12}$, see Eq.~(\ref{J})).

By properly combining the Eq.~(\ref{motion}) for $k=1$, $2$ one is led to
the equation of motion%
\begin{equation}
m_{1}~m_{2}~\frac{d}{dt}\left[ \mathbf{v}_{12}-(\mathbf{I}^{-1}\cdot \mathbf{%
J})\times \mathbf{r}_{12}\right] ~=~(-m_{2}~\mathbf{\nabla }_{1}+m_{1}~%
\mathbf{\nabla }_{2})\left( V(r_{12})+\frac{1}{2}\ \mathbf{J}\cdot \mathbf{I}%
^{-1}\cdot \mathbf{J}\right) ~,
\end{equation}%
where%
\begin{equation}
(-m_{2}~\mathbf{\nabla }_{1}+m_{1}~\mathbf{\nabla }_{2})V(r_{12})~=~\left(
-m_{2}~\frac{\mathbf{r}_{12}}{r_{12}}+m_{1}~\frac{\mathbf{r}_{21}}{r_{12}}%
\right) ~\frac{dV}{dr_{12}}~=~-(m_{1}+m_{2})~\frac{dV}{dr_{12}}~\frac{%
\mathbf{r}_{12}}{r_{12}}~.
\end{equation}%
The gradient of the centrifugal term has been computed in Ref.
\cite{Ferraro}. The
result is%
\begin{equation}
\mu ~\frac{d\mathbf{v}_{12}}{dt}~=~-\frac{dV}{dr_{12}}~\frac{\mathbf{r}_{12}%
}{r_{12}}+2\ \mu \ (\mathbf{I}^{-1}\cdot \mathbf{J})\times \mathbf{v}%
_{12}-\mu \ (\mathbf{I}^{-1}\cdot \mathbf{J})\times \left[ (\mathbf{I}%
^{-1}\cdot \mathbf{J})\times \mathbf{r}_{12}\right] +\mu \ \left[ \frac{d}{dt%
}(\mathbf{I}^{-1}\cdot \mathbf{J})\right] \times \mathbf{r}_{12}~
\label{motion2}
\end{equation}%
($\mu \doteq m_{1}m_{2}/(m_{1}+m_{2})$ is the reduced mass), where
one recognizes the Coriolis, centrifugal and Euler terms associated
not with some \textquotedblleft absolute\textquotedblright\ rotation
but with the intrinsic magnitude $\mathbf{I}^{-1}\cdot \mathbf{J}$
defined by the entire universe. Tensor $\mathbf{I}$ is additive when
the centers-of-mass coincide; so it is $\mathbf{I}=\
\mathbf{I}_{\circledcirc }+\mathbf{I}_{12}$.
$\mathbf{I}_{\circledcirc }$ is an isotropic tensor; besides, one is
free of choosing the $z-$axis along the direction of
$\mathbf{J}_{12}=\mu \ \mathbf{r}_{12}\times \mathbf{v}_{12}$.
Therefore, the tensor $\mathbf{I}$ and its inverse $\mathbf{I}^{-1}$
have the form
\begin{equation}
\mathbf{I}~~=\left(
\begin{array}{ccc}
... & ... & 0 \\
... & ... & 0 \\
0 & 0 & I_{\circledcirc }+\mu \,r_{12}^{2}%
\end{array}%
\right) ~~,\;~\;~\;~\;\mathbf{I}^{-1}~~=\left(
\begin{array}{ccc}
... & ... & 0 \\
... & ... & 0 \\
0 & 0 & (I_{\circledcirc }+\mu \,r_{12}^{2})^{-1}%
\end{array}%
\right) ~~.
\end{equation}%
If one chooses the frame of fixed stars, then it is $\mathbf{J}%
_{\circledcirc }^{\prime}=0$; thus, it follows that%
\begin{equation}
\mathbf{I}^{-1}\cdot \mathbf{J}^{\prime}\ =\ \mathbf{I}^{-1}\cdot
\mathbf{J}_{12}^{\prime}\ =\ \frac{\mathbf{J}_{12}^{\prime}}{%
I_{\circledcirc }+\mu \,r_{12}^{2}}~\doteq ~I^{-1}\ \mathbf{J}%
_{12}^{\prime}~~,
\end{equation}%
which is independent of the choice of the $z-$axis. This result is
substituted in Eq.~(\ref{motion2}) to get%
\begin{equation}
\mu ~\left[ \frac{d\mathbf{v}_{12}^{\prime}}{dt}\right] ^{\prime}~=~-%
\frac{dV}{dr_{12}}~\frac{\mathbf{r}_{12}^{\prime}}{r_{12}}+2\ \mu \
I^{-1}\ \mathbf{J}_{12}^{\prime}\times \mathbf{v}^{\prime}_{12}+\mu
\ I^{-2}\ {J_{12}^{\prime}}^2\ \mathbf{r}_{12}^{\prime}+\mu \ \left[
\frac{d}{dt}(I^{-1}\ \mathbf{J}_{12}^{\prime})\right]
^{\prime}\times \mathbf{r}_{12}^{\prime}~~, \label{motion3}
\end{equation}%
where we used that $\mathbf{J}_{12}^{\prime}$ and $\mathbf{r}%
_{12}^{\prime}$ are mutually perpendicular.

\subsection{Conservation of $I^{-1}\ \mathbf{J}_{12}$ in the frame of fixed
stars}

Let us shows that the Eq.~(\ref{motion3}) leads to the conservation of $%
\mathbf{I}^{-1}\cdot \mathbf{J}_{12}$ in the frame of fixed stars:%
\begin{eqnarray}
\left[ \frac{d}{dt}(I^{-1}\ \mathbf{J}_{12}^{\prime})\right]
^{^{\prime
}}\  &=&\ -I^{-2}\ 2\ \mu \ \mathbf{r}_{12}^{\prime}\cdot \mathbf{v}%
_{12}^{\prime}\ \mathbf{J}_{12}^{\prime}+I^{-1}\ \mu \ \mathbf{r}%
_{12}^{\prime}\times \left[ \frac{d\mathbf{v}_{12}^{\prime}}{dt}%
\right] ^{\prime}  \notag \\
&&  \notag \\
&=&\ -I^{-2}\ 2\ \mu \ \mathbf{r}_{12}^{\prime}\cdot \mathbf{v}%
_{12}^{\prime}\ \mathbf{J}_{12}^{\prime}+I^{-1}\ \mathbf{r}%
_{12}^{\prime}\times \left( 2\ \mu \ I^{-1}\
\mathbf{J}_{12}^{\prime}\times \mathbf{v}_{12}^{\prime}+\mu \ \left[
\frac{d}{dt}(I^{-1}\
\mathbf{J}_{12}^{\prime})\right] ^{\prime}\times \mathbf{r}%
_{12}^{\prime}\right)   \notag \\
&&  \notag \\
&=&\ I^{-1}\ \mu \ \mathbf{r}_{12}^{\prime}\times \left( \left[ \frac{d}{%
dt}(I^{-1}\ \mathbf{J}_{12}^{\prime})\right] ^{\prime}\times \mathbf{%
r}_{12}^{\prime}\right)
\end{eqnarray}%
Therefore one obtains
\begin{equation}
\left[ \frac{d}{dt}(I^{-1}\ \mathbf{J}_{12}^{\prime})\right]
^{\prime}\ =\ 0\ .
\end{equation}%
The conservation of $I^{-1}\ \mathbf{J}_{12}^{\prime}$ reduces the
equations of motion to the form%
\begin{equation}
\mu ~\left[ \frac{d\mathbf{v}_{12}^{\prime}}{dt}\right] ^{\prime}~=~-%
\frac{dV}{dr_{12}}~\frac{\mathbf{r}_{12}^{\prime}}{r_{12}}+2\ \mu \
I^{-1}\ \mathbf{J}_{12}^{\prime}\times \mathbf{v}^{\prime}_{12}+\mu
\ I^{-2}\ {J_{12}^{\prime}}^2\ \mathbf{r}_{12}^{\prime}\ .
\label{motion4}
\end{equation}%
These equations are formally identical to those of Newton's
dynamics, as regarded from a (non-inertial) frame that rotates with
(absolute) constant velocity $\mathbf{\Omega }=-$ $I^{-1}\
\mathbf{J}_{12}^{\prime}$. Although the frame of fixed stars is
non-Newtonian, the relational dynamics of the studied subsystem can
be easily recovered from Newtonian dynamics by
means of a simple correction involving $\mathbf{\Omega }$. However, $\mathbf{%
\Omega }$ is not an absolute magnitude but it is directly related to
the initial conditions characterizing the solution\footnote{The
conservation of $I^{-1}\ \mathbf{J}^{\prime} _{12}$ can be regarded
as a consequence of the conservation of $\mathbf{J}_{12}$ in the
Newtonian frame
and the way $\mathbf{J}_{12}$ transforms under change of frame (cf.~Eq.~(\ref%
{Jprime}))\label{fn8}.}.

\subsection{Energy conservation in the frame of fixed stars}

To study the energy conservation we will integrate the scalar
multiplication of Eq.~(\ref{motion4}) with
$\mathbf{v}_{12}^{^{\prime
}}\ dt=d\mathbf{r}_{12}^{\prime}$. Thus, it follows that%
\begin{equation}
\frac{1}{2}\ \mu \
|\mathbf{v}_{12}^{\prime}|^{2}+V(r_{12})-\frac{1}{2}\ \mu \ I^{-2}\
{J_{12}^{\prime}}^2\ r_{12}^{2}\ =\ \text{constant}\ .
\end{equation}%
We normally use this equation to describe the radial motion, by using the
decomposition%
\begin{equation}
|\mathbf{v}_{12}^{\prime }|^{2}\ =\ v_{radial}^{2}+v^{\prime\
2}_{tangential}\ =\ v_{radial}^{2}+\frac{{J_{12}^{\prime}}^2}{\mu
^{2}\ r_{12}^{2}}\ .
\end{equation}%
Thus, the effective potential for the radial motion is%
\begin{equation}
V_{eff}\ =\ V(r_{12})+\frac{(I^{-1}\ J_{12}^{\prime})^{2}}{2\ \mu \
r_{12}^{2}}\ \left( I^{2}-\mu ^{2}\ r_{12}^{4}\right) \ =\ V(r_{12})+\frac{%
(I^{-1}\ J_{12}^{\prime})^{2}\ I_{\circledcirc }^{2}}{2\ \mu \ r_{12}^{2}%
}+(I^{-1}\ J_{12}^{\prime})^{2}\ I_{\circledcirc }\ .\label{eff}
\end{equation}%
The radial motion is gauge invariant, since the distance between
particles is an observable. This fact reflects in the effective
potential (\ref{eff}), which keeps the form of its Newtonian version
(notice that, according to Footnote \ref{fn8}, the conserved
quantity $I_{\circledcirc }\ I^{-1}\ J_{12}^{\prime}$ is equal to
the value of $J_{12}$ in the Newtonian frame).

\section{\protect\bigskip Conclusions}

\label{SVI}Newton's mechanics governs the evolutions of individual
particles in the absolute space. Particle positions express
themselves in Newton's laws by means of coordinates referred to a
frame at rest or in uniform translation with respect to the absolute
space (Galileo's symmetry). Although Galileo's symmetry implies that
absolute motion is undetectable, it confers the acceleration the
status of an absolute property (independent of the chosen inertial
frame). Instead, relational mechanics is a theory that governs the
dynamics of the distances between particles. Distances do not
require a frame to manifest themselves. For practical reasons, we
still use a frame to write distances in terms of particle
coordinates. But the description of a configuration in terms of
distances and their derivatives is completely frame-independent.
Thus the (time-dependent) changes of frames constitute a gauge
symmetry in relational mechanics. Therefore, the idea of absolute
space as an entity that selects the allowed frames is vain in
relational mechanics, since frames are reduced to the role of useful
accessories. These two approaches to the laws of mechanics are
conceptually very different. If a subsystem is isolated, in the
sense that its interaction with particles outside the subsystem is
negligible, Newton's laws describe its evolution just in terms of
the (absolute) initial conditions of its own particles. Instead,
relational mechanics always describes a subsystem in terms of
internal and external distances; so, even if the subsystem is
\textquotedblleft isolated\textquotedblright , its evolution will
anyway depend on the relation between the subsystem and the rest of
the universe. This is so because $\mathbf{I}^{-1}\cdot \mathbf{J}$
takes part in all the equations of motion; $\mathbf{I}^{-1}\cdot
\mathbf{J}$ is the essential piece to guarantee the gauge invariance
of the relational dynamics. However, we can exploit the gauge
invariance by choosing a frame where the intrinsic angular momentum
of the universe vanishes at each instant. In such frames, which are
defined by the entire universe, the equations of motion become the
Newton's laws (Newtonian frames). Newton's laws reappear as
gauge-fixed equations of motion because the relational Lagrangian
was defined by gauging the Newtonian Lagrangian. This way of
recovering the Newtonian dynamics implies that a Newtonian solution
for an isolated subsystem is valid if and only if it is part of a
Newtonian solution with $\mathbf{J}=0$ for the entire universe. In
particular, a spinning Newtonian solution for an isolated subsystem
can only work if there exists a rest of universe to make feasible
the condition $\mathbf{J}=0$. With the aim of analyzing the
consequences of this statement, we have proposed a very simplified
--non realistic-- model where the rest of the universe is
represented by a rigid isotropic shell centered at the
center-of-mass of the subsystem under consideration. Thus, the
subsystem remains gravitationally isolated, and $\mathbf{J}$ can be
decomposed as the sum of the intrinsic angular momenta of the shell
and the subsystem (see Footnote \ref{fn}). This simple arrangement
facilitates the building of Newtonian solutions for the entire
universe; in fact, the (conserved) angular momentum of the Newtonian
solution for the isolated subsystem is trivially compensated by the
shell (Newtonianly) rotating at a constant velocity. In this crude
model, the frame of fixed stars is the frame where the shell is at
rest\footnote{The International Celestial Reference Frame (ICRF2) is
defined by the positions of about 300 extragalactic sources.}; this
is not a Newtonian frame since $\mathbf{J}$ is not null but is equal
to the subsystem intrinsic angular momentum
$\mathbf{J}_{S}^{\prime}$. Then, we cannot expect to observe the
Newtonian solution in the frame where the shell is at rest;\ instead
we will observe that the Newtonian solution is dragged by a rotation
$\mathbf{\Omega }=-$ $I^{-1}\ \mathbf{J}_{S}^{\prime}\simeq -$
$I_{\circledcirc }^{-1}\ \mathbf{J}_{S}^{\prime}$. As seen, the
larger is $\mathbf{J}_{S}^{\prime}$, the greater is the dragging
effect (with $I_{\circledcirc }$ playing the role of a universal
constant). As we have shown, the dragging effect alters the galactic
rotation curves and the two-body dynamics when observed from the
frame where the shell is at rest. Consequently, the virial theorem
also gets terms associated with $\mathbf{J}_{S}^{\prime}$ (see the
Eq.~(\ref{approxvirial}) for the simpler approximate result).
Notoriously, these effects of relational mechanics take part in the
phenomena that led to the hypothesis of dark matter: the discrepancy
between luminous masses in galaxies and clusters of galaxies and the
masses inferred from the virial theorem \cite{Zwicky33,Zwicky37},
and the dynamics in galactic halos \cite{Rubin, Persic}. Even so, it
should be noticed that we have not estimated their contributions to
such phenomena, which depend on the intrinsic angular momenta of
galaxies or clusters of galaxies, and the unknown ``universal
constant'' $I_{\circledcirc }$. Anyway, it is worth mentioning that
the relational dragging effect can be separated from dark matter
effects: while dark matter acts in the same way whatever the
direction of the rotation is, the dragging effect increases the
velocities of co-rotating objects but decreases the velocities of
the counter-rotating ones, since it depends on the sign of
${J}_{\Game }$ in Eq.~(\ref{Coriolis}).

\bigskip

\begin{acknowledgments}
This work was supported by Consejo Nacional de Investigaciones
Cient\'{\i}ficas (CONICET) y T\'{e}cnicas and Universidad de Buenos Aires.
\end{acknowledgments}


\begin{thebibliography}{99}
\bibitem{Mach} Mach, E.: Die Mechanik in ihrer Entwicklung.
Historisch-kritisch dargestellt. F.A. Brockhaus, Leipzig (1883) (The
Science of Mechanics: A Critical and Historical Account of Its
Development. The Open Court Publishing Co., Chicago (1893))

\bibitem{Alexander} Alexander, H.G.: The Leibniz-Clarke
Correspondence Together with Extracts from Newton's Principia and
Opticks. Manchester University Press, Manchester (1956)

\bibitem{Ferraro} Ferraro, R.: Relational Mechanics as a gauge theory, Gen. Relat. Gravit. \textbf{48}, 23 (2016)

\bibitem{Einstein12} Einstein, A.: Gibt es eine Gravitationswirkung,
die der elektrodynamischen Induktionswirkung analog ist?,
Vierteljahrsschrift f\"{u}r gerichtliche Medizin und \"{o}ffentliches
Sanit\"{a}tswesen \textbf{44}, 37-40 (1912)

\bibitem{Einstein14} Einstein, A.: Die formale Grundlage der allgemeinen Relativit\"{a}tstheorie, Sber. Preuss. Akad. Wiss. Berlin,
1030-1085 (1914)

\bibitem{Einstein16} Einstein, A.: Die Grundlage der allgemeinen Relativit\"{a}tstheorie, Annalen der Physik
\textbf{354}, 769-822 (1916)

\bibitem{Einstein18} Einstein, A.: Prinzipielles zur allgemeinen Relativit\"{a}tstheorie, Annalen der Physik \textbf{360},
241-244 (1918)

\bibitem{Hofmann} Hofmann, W.: Kritische Beleuchtung der beiden
Grundbegriffe der Mechanik: Bewegung und Tr\"{a}gheit und daraus
gezogene Folgerungen betreffs der Achsendrehung der Erde und des
Foucault'schen Pendelversuchs. M. Kuppitsch Wwe., Wien (1904)
(partial English translation in Ref.~\onlinecite{Barbour95})

\bibitem{Reissner} Reissner, H.: \"{U}ber die Relativit\"{a}t der Beschleunigungen in der Mechanik, Physikalische Zeitschrift \textbf{15},
371-375 (1914) (English translation in Ref.~\onlinecite{Barbour95})

\bibitem{Schrodinger} Schr\"{o}dinger, E.: Die Erf\"{u}llbarkeit der Relativit\"{a}tsforderung in der klassischen Mechanik, Annalen der Physik \textbf{382}, 325-336 (1925) (English translation in
Ref.~\onlinecite{Barbour95})

\bibitem{Barbour75} Barbour, J.B.: Forceless machian dynamics, Nuovo Cimento B \textbf{26}, 16-22 (1975)

\bibitem{Barbour95} Barbour, J.B. and Pfister, H. (eds.): Einstein
Studies, vol. 6: Mach's Principle: From Newton's Bucket to Quantum
Gravity. Birkh\"{a}user, Boston (1995)

\bibitem{Hughes} Hughes, V.W., Robinson, H.G.  and Beltran-Lopez, V.: Upper limit for the anisotropy of inertial mass from nuclear resonance experiments, Phys. Rev.
Lett. \textbf{4}, 342-344 (1960)

\bibitem{Barbour77} Barbour, J.B. and Bertotti, B.: Gravity and Inertia in a Machian Framework, Nuovo Cimento B \textbf{38}, 1-27 (1977)

\bibitem{Barbour82} Barbour, J.B. and Bertotti, B.: Mach's principle and the structure of dynamical theories, Proceedings of the Royal Society London A
\textbf{382}, 295-306 (1982)

\bibitem{Barbour03} Barbour, J.: Scale-invariant gravity: particle dynamics, Classical Quant. Grav. \textbf{20},
1543-1570 (2003)

\bibitem{Gryb} Gryb, S.: Implementing Mach's principle using gauge theories, Phys. Rev. D \textbf{80}, 024018 (2009)

\bibitem{Anderson} Anderson, E.: The problem of time and quantum cosmology in the relational particle mechanics arena. arXiv:1111.1472v3 (2011)

\bibitem{Mercati} Mercati, F.:  A Shape Dynamics Tutorial. arXiv:1409.0105 (2014)

\bibitem{Lynden92} Lynden-Bell, D.:  The relativity of acceleration. In: Warner, B. (ed.) Variable
Stars and Galaxies (in honour of M.W. Feast), ASP Conference Series
\textbf{30} (1992)

\bibitem{Lynden95} Lynden-Bell, D.:  A Relative Newtonian Mechanics. In: Barbour, J.B. and Pfister, H. (eds.)
Einstein Studies, vol. 6: Mach's Principle: From Newton's Bucket to
Quantum Gravity. Birkh\"{a}user, Boston (1995)

\bibitem{Katz95} Lynden-Bell, D. and Katz, J.:  Classical mechanics without absolute space, Phys. Rev. D \textbf{52}, 7322-7324 (1995)

\bibitem{Zwicky33} Zwicky, F.: Die Rotverschiebung von extragalaktischen Nebeln, Helv. Phys. Acta \textbf{6}, 110-127 (1933)

\bibitem{Zwicky37} Zwicky, F.:  On the masses of nebulae and of clusters of nebulae, Astrophys. J. \textbf{86}, 217-246 (1937)

\bibitem{Rubin} Rubin, V.C.,  Ford, W.K. and Thonnard, N.: Rotational properties of 21 Sc galaxies with a large range
of luminosities and radii, from NGC 4605 (R = 4 kpc) to UGC 2885 (R
= 122 kpc), Astrophys. J. \textbf{238}, 471-487 (1980)

\bibitem{Persic} Persic, M.,  Salucci, P. and Stel, F.:  The universal rotation curve of spiral galaxies-I. The dark matter connection, Mon. Not. R. Astron.
Soc. \textbf{281}, 27-47 (1996)
\end{thebibliography}
\end{document}